\documentclass[a4paper, 11pt]{article}
\usepackage{comment}
\usepackage{lipsum}
\usepackage{fullpage}
\usepackage{graphicx, setspace}
\usepackage{caption}
\usepackage{subcaption}
\usepackage{newfloat}
\usepackage{rotating}
\usepackage{pgfplots}
\usepackage{amsmath}
\usepackage{listings}
\usepackage{placeins}
\usepackage{hyperref}
\usepackage{epstopdf}
\DeclareFloatingEnvironment[%
	fileext=los,
	name=Exhibit,
]{exhibit}

\begin{document}
\noindent
\begin{center}
\LARGE\textbf{Rational Decision-Making Under Uncertainty: 
Observed Betting Patterns on a Biased Coin}\footnote{The authors are grateful for the help of our colleagues, friends and family members who helped us design and program the experiment, test it out, and then put our findings into this note. In particular, thank you to Samantha McBride, Adam Smith, Jinwoo Baek, Larry Hilibrand, Vlad Ragulin, William Ziemba, Edward O. Thorp, Andrew Morton, Chris Rokos, Jamil Baz, Rob Stavis, Jeff Rosenbluth, James W. White, Jessica Haghani and Joshua Haghani.} \\\hspace{1cm} 

\large{Victor Haghani\footnote{Victor is Founder and CEO of Elm Partners, victor@elmfunds.com}\\
Richard Dewey\footnote{Rich works at PIMCO, rld2126@columbia.edu}}\\
\large{Working Draft} \\October 19, 2016 \\

\end{center}
\section{Introduction}
You$'$re invited to a talk by a hedge fund manager who was a partner at a fund that famously flopped about twenty years ago. You turn up, hoping to hear some valuable insights, or at least some entertaining tales, but instead you are offered a stake of \$25 to take out your laptop to bet on the flip of a coin for thirty minutes. You$'$re told the coin is biased to come up heads with a 60\% probability, and you can bet as much as you like on heads or tails on each flip. You will be given a check for however much is in your account at the end of the half hour.\footnote{Subject to a maximum payout that you$'$ll be informed of if you get close.}  

That$'$s it. Would you feel it was worth your time to play, or would you walk out? How would you play the game? What heuristic or mental tool-kit would you employ? These questions led us to conducting the exact experiment described above. By having participants engage in an activity as simple as flipping a coin, the betting strategy and its evolution are easily isolated for observation. This simple game also turns out to have properties that are similar to investing in the stock market as well as implications for finance and economics education. 

Below we'll describe the experiment, how our subjects played the game and the conclusions we draw from the experiment. 

\section{The Experiment}

Our coin-flipping experiment was played by 61 subjects, in groups of 2-15, in the quiet setting of office conference rooms or university classrooms. The proctor for the game outlined basic principles, such as no talking or cooperation and that subjects were not to use the internet or other resources while playing the game. 

The experiment began when subjects were directed to a URL that contained a purpose-built application for placing bets on the flip of a simulated coin. Participants used their personal laptops or work computers to play the game. Prior to starting the game, participants read a detailed description of the game, which included a clear statement, in bold, indicating that the simulated coin had a 60\% chance of coming up heads and a 40\% chance of coming up tails. Participants were given \$25 of starting capital and it was explained in text and verbally that they would be paid, by check, the amount of their ending balance subject to a maximum payout. The maximum payout would be revealed if and when subjects placed a bet that if successful would make their balance greater than or equal to the cap. We set the cap at \$250, ten times the initial stake. Participants were told that they could play the game for thirty minutes, and if they accepted the \$25 stake, they had to remain in the room for that amount of time.\footnote{Whether they chose to not play, or did play and went bust or hit the cap.} Participants could place a wager of any amount in their account, in increments of \$0.01, and they could bet on heads or tails. Participants were asked a series of questions about their background before playing and about their experience when they finished. 

The sample was largely comprised of college age students in economics and finance and young professionals at finance firms. We had 14 analyst and associate level employees at two leading asset management firms. The sample consisted of 49 males and 12 females. Our prior was that these participants should have been well prepared to play a simple game with a defined positive expected value.

\section{Optimal Strategy}

Before continuing with a description of what an optimal strategy might look like, perhaps you$'$d like to take a few moments to consider what you would do if given the opportunity to play this game. Once you read on, you$'$ll be afflicted with the curse of knowledge, making it difficult for you to appreciate the perspective of our subjects encountering this game for the first time. So, if you want to take a moment to think about your strategy, this is the time to do it. 

If you$'$re a professional gambler, chances are you$'$ve heard of the Kelly criterion, a formula published in 1955 by John Kelly, a brilliant if somewhat eccentric researcher working at Bell Labs. The formula provides an optimal betting strategy for maximizing the rate of growth of wealth in games with favorable odds, a tool that would appear a good fit for this problem. Dr. Kelly$'$s paper built upon work first done by Daniel Bernoulli, who resolved the St. Petersburg Paradox - a lottery with an infinite expected payout -  by introducing a utility function that the lottery player seeks to maximize. Bernoulli$'$s work catalyzed the development of utility theory and laid the groundwork for many aspects of modern finance and behavioral economics. 

Dr. Kelly$'$s paper and the eponymous formula, caught the attention of gamblers and investors. It was further developed and applied to casino games and financial markets by Ed Thorp in a series of papers and popular books, most notably Beat the Dealer and Beat the Market. Following Kelly and Thorp$'$s initial work, many others, including Murray Gell-Mann have further developed the theoretical foundations, while notable investors such as Warren Buffett, Bill Gross and James Simons have all reportedly made use of the Kelly formula. 

The basic idea of the Kelly formula is that a player who wants to maximize the rate of growth of his wealth should bet a constant fraction of his wealth on each flip of the coin, defined by the function 2*p-1, where p is the probability of winning. The formula implicitly assumes the gambler has log utility. It$'$s intuitive that there should be an optimal fraction to bet; if the player bets a very high fraction, he risks losing so much money on a bad run that he would not be able to recover, and if he bet too little, he would not be making the most of what is a finite opportunity to place bets at favorable odds. While it$'$s true that the expected value of the game goes up the higher the fraction the player bets, the outcomes become so skewed that a player who exhibits risk aversion will find an optimal betting fraction well below 100\%. The odds themselves play a role in the optimal fraction to bet; the more favorable the odds, the higher a fraction one ought to bet. Finally, as the flips are independent random outcomes, the strategy should only depend on the player$'$s account balance, and not on the pattern of previous flips.\footnote{We present the Kelly criterion as a useful heuristic a subject could gainfully employ. It may not be the optimal approach for playing the game we presented for several reasons. The Kelly criterion is consistent with the bettor having log-utility of wealth, which is a more tolerant level of risk aversion than most people exhibit. On the other hand, the subjects of our experiment likely did not view \$25 (or even \$250) as the totality of their capital, and so they ought to be less risk averse in their approach to maximizing their harvest from the game. The fact that there is some cap on the amount the subject can win should also modify the optimal strategy.}

In our game, the Kelly criterion would tell the subject to bet 20\% (2*.6-1) of his account on heads on each flip. So, the first bet would be \$5 (20\% of \$25) on heads, and if he won, then he$'$d bet \$6 on heads (20\% of \$30), but if he lost, he$'$d bet \$4 on heads (20\% of \$20), and so on.

\section{Findings: How Well Did Our Players Play?}
\textit{How did you go bankrupt? Gradually, and then suddenly.}\ (Ernest Hemingway, The Sun Also Rises 1926) \\

Our subjects did not do very well. While we expected to observe some suboptimal play, we were surprised by the pervasiveness of it. Suboptimal betting came in all shapes and sizes: over-betting, under-betting, erratic betting and betting on tails were just some of the ways a majority of players squandered their chance to take home \$250 for 30 minutes play. 

Only 21\% of participants reached the maximum payout of \$250,\footnote{We define $'$maxing out$'$ as players who reached at least \$200 by the end, and we define $'$going bust$'$ as those finishing the game with less than \$2 in their account at the end.} well below the 95\% that should have reached it given a simple constant percentage betting strategy of anywhere from 10\% to 20\%.\footnote{A result we calculated through Monte Carlo simulation. See Section 5 for more detail.} 

We were surprised that one third of the participants wound up with less money in their account than they started with. More astounding still is the fact that 28\% of participants went bust and received no payout. That a game of flipping coins with an ex-ante 60/40 winning probability produced so many subjects that lost everything is startling.

The average ending bankroll of those who did not reach the maximum and who also did not go bust, which represented 51\% of the sample, was \$75. While this was a tripling of their initial \$25 stake, it still represents a very sub-optimal outcome given the opportunity presented. The average payout across all subjects was \$91, letting the authors off the hook relative to the \$250 per person they$'$d have had to pay out had all the subjects played well. The chart below summarizes the performance of our 61 subjects, who in aggregate wagered on 7,253 coin flips, 59.6\% of which were heads. 

\begin{exhibit}[h!]
\centering
\includegraphics[width=1.0\textwidth]{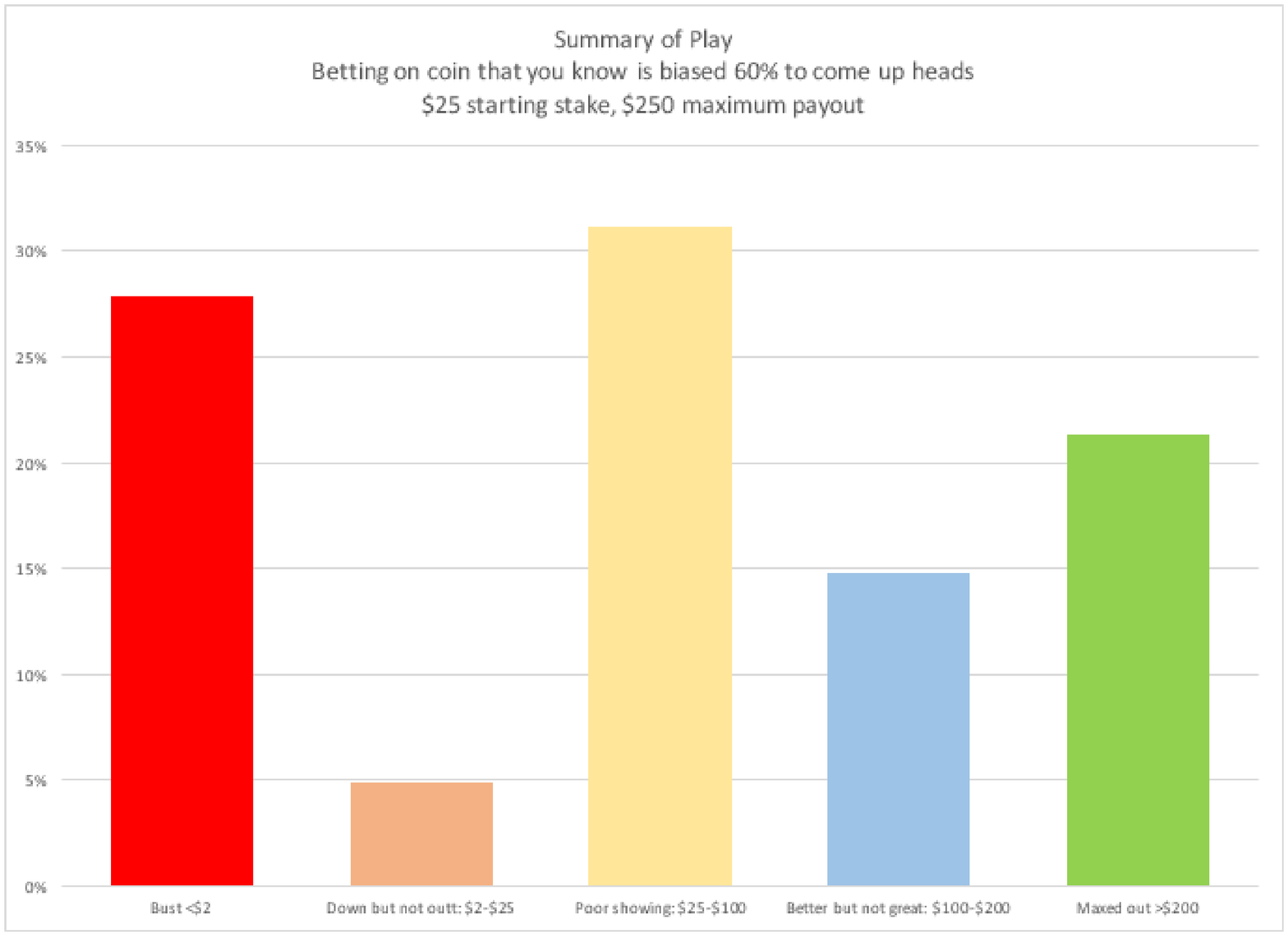}
\caption{coin flip performance}
\end{exhibit}

Only 5 of our 61 financially sophisticated students and young investment professionals reported that they had ever heard of the Kelly criterion. Interestingly, having heard of Kelly did not seem to help two of them: one barely managed to double his stake, and the other one only broke even after about 100 flips. In post-experiment interviews, we found that the notion of betting a constant proportion of wealth seemed to be a surprisingly non-intuitive approach to playing this game. Our results do not offer any indication that participants were converging to optimal play over time as evidenced by suboptimal betting of similar magnitude throughout the game. 

How subjects played the game in the absence of employing Kelly was illuminating. Of the 61 subjects, 18 subjects bet their entire bankroll on one flip, which increased the probability of ruin from close to 0\% using Kelly to 40\% if their all-in flip was on heads, or 60\% if they bet it all on tails, which amazingly some of them did. The average bet size across all subjects was 15\% of the bankroll, so participants bet less, on average, than the Kelly criterion fraction, which would make sense in the presence of a maximum payout that would be within reach. However, this apparent conservatism was completely undone by participants generally being very erratic with their fractional betting patterns, betting too small and then too big. Betting patterns and post experiment interviews revealed that quite a few participants felt that some sort of doubling down, or Martingale betting strategy, was optimal, wherein the gambler increases the size of his wagers after losses. Another approach followed by a number of subjects was to bet small and constant wagers, apparently trying to reduce the probability of ruin and maximize the probability of ending up a winner.

We observed 41 subjects (67\%) betting on tails at some point during the experiment. Betting on tails once or twice could potentially be attributed to curiosity about the game, but 29 players (48\%) bet on tails more than 5 times in the game. It is possible that some of these subjects questioned whether the coin truly had a 60\% bias towards heads, but that hypothesis is not supported by the fact that within the subset of 13 subjects who bet on tails more than 25\% of the time, we found they were more likely to make that bet right after the arrival of a string of heads. This leads us to believe that some combination of the illusion of control, law of small numbers bias, gamblers fallacy or hot hand fallacy was at work. After the game concluded, we asked participants a series of questions, including whether they believed the coin actually had a 60\% bias towards heads. Of those who answered that question, 75\% believed that was the case. You can see all the pre and post-trial questions by playing the game (we won$'$t pay you though) here: 
\url{ http://coinflipbet.herokuapp.com/}

\section{How Much Should You Be Willing to Pay to Play?}

Not only did most of our subjects play poorly, they also failed to appreciate the value of the opportunity to play the game. 
If we had offered the game with no cap, this experiment could have become very, very expensive for your authors.\footnote{In fact, one reason we suspect this experiment was not performed until now is that it is quite an expensive undertaking even with just 60 subjects.} Assuming a player with agile fingers can put down a bet every 6 seconds, that would allow 300 bets in the 30 minutes of play.\footnote{We programmed the coin to be in a flipping mode for about 4 seconds, to create some suspense on each flip, and also to limit the number of flips to about 300.} The expected gain of each flip, betting the Kelly fraction, is 4\%\ and so the expected value of 300 flips is \$25 * $(1 + .04)^{300}$ = \$3,220,637! \footnote{ Your opening bet, according to Kelly, would be \$5 on heads. The expected gain from that flip would be \$1, as there is a 60\% chance of winning \$5 and a 40\% chance of losing \$5 = 0.6*5-0.4*5 = \$1. Your capital in the game at the moment you place that bet is \$25, so the expected return on capital is 4\% (\$1/\$25).  Each successive flip of the coin will have that same 4\% expected return, up until the cap is encountered, or if the subject gets down to \$0.04 or less, at which point he cannot bet 20\% of his account any more as we limit the subject to betting \$0.01 or more on each flip. And that$'$s just the expected value. If a subject was very lucky, and flipped 210 heads and only 90 tails (admittedly very unlikely), then we$'$d have owed him about \$2 billion!}

Given the expected value of the uncapped game is about \$3 million, how much should a person be willing to pay to play this game, assuming that he believes that the person offering the game has enough money to meet all possible payouts?\footnote{Of course, this is not realistic as that would be about \$14 trillion trillion (\$25 * $1.2^{300}$). We suspect that not even the Fed, ECB and BoJ working together could print that much money.} Just as is the case with the St Petersburg Paradox, where players are generally unwilling to pay more than \$10 to play a game with an infinite expected value, in our game too, players should only be willing to pay a fraction of the \$3 million expected value of the game.\footnote{ As with the St Petersburg paradox, much of the high expected value of our game comes from unlikely but very big positive outcomes. The skew can also be seen from the fact that the median of the distribution is so much lower than the mean, which arises from the fact that if you bet 20\% of your account and win, you go up to 1.2 of your wealth, and then if you bet 20\% of that and lose, you now wind up at 1.2*0.8=0.96, or 4\% less than what you had. The median outcome of 180 heads (0.6*300) and 120 tails would produce an outcome of only \$10,504 (\$25*$1.2^{180}$ * $0.8^{120}$), much below the \$3,220,637 expected value.} For example, if we assume our gambler has log utility (which the Kelly solution implies) and has de minimus investable wealth, then he should be willing to pay about \$10,000 to play the game (the dollar equivalent of the expected utility), a small fraction of \$3 million, but still a very large absolute amount of money in light of the \$25 starting stake.\footnote{ For each flip, the expected utility is 0.6*ln(1.2)+0.4*ln(0.8)= 0.0201, and exp(0.0201)=1.02034, which means that each flip is giving a dollar equivalent increase in utility of about 2\% and so for 300 flips, we get \$25*$1.02034^{300}$ = \$10,504. This is also the median of the distribution, as per above footnote. If we relax the assumption regarding the player having no outside wealth, the amount he should be willing pay can be much higher than \$10,000.}

With a capped payout (the game we actually offered) a simple (but not strictly optimal) strategy would incorporate an estimate of the maximum payout. If the subject rightly assumed we wouldn$'$t be offering a cap of more than \$1,000 per player, then a reasonable heuristic would be to bet a constant proportion of one$'$s bank using a fraction less than the Kelly criterion, and if and when the cap is discovered, reducing the betting fraction further depending on betting time remaining to glide in safely to the maximum payout. For example, betting 10\% or 15\% of one$'$s account may have been a sound starting strategy. 

We ran simulations on the probability of hitting the cap if the subject bet a fixed proportion of wealth of 10\%, 15\% and 20\%, and stopping when the cap was exceeded with a successful bet. We found there to be a 95\% probability that the subjects would reach the \$250 cap following any of those constant proportion betting strategies, and so the expected value of the game as it was presented (with the \$250 cap) would be just under \$240. However, if they bet 5\% or 40\% of their bank on each flip, the probability of exceeding the cap goes down to about 70\%. 

\section{Similarities to Investing in the Stock Market}

\textit{If you gave an investor the next day$'$s news 24 hours in advance, he would go bust in less than a year.}\ (Nassim Taleb) \\

An interesting aspect of this experiment is that it has significant similarities to investing in the stock market. For example, the real return of US equities over the past 50 years was a bit over 5\% and the annual standard deviation was about 15\%, giving a return/risk ratio of about 0.33. Many market observers believe the prospective return/risk ratio of the stock market is well below its historical average, and closer to that of our coin flip opportunity, which has a return/risk ratio of 0.2.\footnote{More precisely the ratio is 0.204}

%\begin{exhibit}[h!]
%\centering
%\includegraphics[width=1.0\textwidth]{/Users/rdewey/Desktop/flip.eps}
%\includegraphics[width=0.9\textwidth]{coinflippaul.jpeg}
%\caption{Picture of guy flipping a coin}
%\end{exhibit}

Of course, there are significant differences, from the binary versus continuous nature of outcomes, to the question of risk versus uncertainty when investing in the stock market where no one can tell you the distribution from which you will draw outcomes. Furthermore, most investors believe the stock market is not a successive set of independent flips of a coin, but that there are elements of mean reversion and trending in stock market behavior, and of course, outlier events happen with much higher probability than would evolve from a series of coin flips.\footnote{Perhaps more nuanced, our coin flip game generates a distribution where you make or lose a fixed amount on each flip, whereas many people believe the stock market has more of a lognormal distribution where the positive flip outcome is greater than the loss from a negative flip. That is, stocks may be characterized by outcomes of $e^{d}$ and $e^{-d}$, whereas our coin flip has 1+d and 1-d for outcomes.}

Most people we discussed this with felt that there is a fundamental difference between flipping a coin 300 times in 30 minutes, and investing in the stock market where we have to wait 30 years to get 30 flips of the coin. In fact, to the extent that stocks follow a random walk, with both return and the risk we care about, variance, both growing proportionately with time, then horizon should not affect our betting strategy, although it does affect how highly we value the opportunity to play.\footnote{There are a number of assumptions in this statement, including that we display constant relative risk aversion, a common but certainly not the only representation of risk aversion among classic and modern behavioral models.}

After the experiment, we discussed Kelly and optimal betting strategies with our subjects. We were left with the feeling that they would play the game more effectively if given another chance. We wonder whether any long lasting impact could be had on investor behavior through similar discussions of sensible approaches to stock market investing. Perhaps investing in the stock market is much more nuanced and complex than betting on a biased coin, or perhaps it$'$ easy to stick to a sound, albeit boring, strategy for 30 minutes but impossible to maintain that discipline for 30 weeks, months or years.

\section{Conclusion}

\textit{This is a great experiment for many reasons. It ought to become part of the basic education of anyone interested in finance or gambling}\ (Edward O. Thorp) \\

While we did expect to observe poorly conceived betting strategies from our subjects, we were surprised by the fact that 28\% of our subjects went bust betting on a coin that they were told was biased to come up heads 60\% of the time. Before this experiment, we did not appreciate just how ill-equipped so many people are to appreciate or take advantage of a simple advantageous opportunity in the presence of uncertainty. The straightforward notion of taking a constant and moderate amount of risk and letting the odds work in one$'$s favor just doesn$'$t seem obvious to most people.  

Given that many of our subjects received formal training in finance, we were surprised that the Kelly Criterion was virtually unknown and that they didn$'$t seem to possess the analytical tool-kit to lead them to constant proportion betting as an intuitively appealing heuristic. Without a Kelly-like framework to rely upon, we found that our subjects exhibited a menu of widely documented behavioral biases such as illusion of control, anchoring, over-betting, sunk-cost bias, and gambler$'$s fallacy. We reviewed the syllabi of introductory finance courses and elective classes focused on trading and asset pricing at five leading business schools in the United States.\footnote{MIT, Columbia, Chicago, Stanford and Wharton.} Kelly was not mentioned in any of them, either explicitly, or by way of the topic of optimal betting strategies in the presence of favorable odds. Could the absence of Kelly be the effect of Paul Samuelson$'$s vocal critique of Kelly in public debate with Ed Thorp and William Ziemba?\footnote{Ziemba (2016)}. If so, it$'$s time to bury the hatchet and move forward. 

These results raise important questions. If a high fraction of quantitatively sophisticated, financially trained individuals have so much difficulty in playing a simple game with a biased coin, what should we expect when it comes to the more complex and long-term task of investing one$'$s savings? Is it any surprise that people will pay for patently useless advice, as documented in studies like Powdthavee (2012)? What do the results of this experiment say about the prospects for reducing wealth inequality, or ensuring the stability of our financial system? 

Our research suggests there is a meaningful gap in the education of young finance and economics students when it comes to the practical application of the concepts of utility and risk taking. The existence of this gap is even more surprising than the poor play of our subjects. After all, can we really blame them if they haven$'$t received sufficient practical training? Our research will be worth many multiples of the \$5,574 winnings we paid out to our 61 subjects if it helps encourage educators to fill this void, either directly through more instruction or through trial and error exercises like our game.

\newpage

\section*{References}  

\begin{itemize}

\item Choi, James, David Laibson, and Brigitte Madrian. 2010. Why does the law of one price fail? An experiment on index mutual funds. Review of Financial Studies 23(4): 1405-1432.

\item Fenton-O-Creevy, Mark; Nicholson, Nigel; Soane, Emma and Willman, Paul. 2003. Trading on Illusions: Unrealistic perceptions of control and trading performance. 

\item Friedland, Keinan and Regev. 1992. Controlling the Uncontrollable: Effects of Stress on Illusory Perceptions of Controllability. 

\item Gilovich, Thomas; Tversky, A.; Vallone, R. (1985). The Hot Hand in Basketball: On the Misperception of Random Sequences. Cognitive Psychology 3.

\item Green, Brett; Zwiebel, Jeffery. The Hot Hand Fallacy: Cognitive Mistakes or Equilibrium Adjustments? Evidence from Baseball. Stanford Graduate School of Business. Retrieved 2016-05-06.

\item Langer, Ellen J. 1975. The Illusion of Control. The Journal of Personality and Social Pyschology. 

\item Levitt, Steven. 2016. Head or Tails: The Impact of a Coin Toss on Major Life Decision and Subsequent Happiness. NBER working paper.

\item Kelly, J.L., 1956. A new interpretation of information rate. Bell System Technical Journal 35, 917-926. 

\item MacLean, Thorp and Ziemba editors. "The Kelly Capital Growth Investment Criterion," World Scientific 2010. 

\item Miller, Joshua; Sanjurjo, Adam. 2015. A Cold Shower for the Hot Hand Fallacy.

\item Powdthavee, Nattavudh and Riyanto, Yohanes, E., 2012 Why Do People Pay for Useless Advice? Implications of Gamblers and Hot-Hand Fallacies in False-Expert Setting. Working Paper (IZA DP No. 6557)

\item Rotando, L.M., Thorp, E.O., 1992. The Kelly criterion and the stock market. American Mathematical Monthly, 922?931. 

\item Taleb, Nassim N. 2016. Mathematical Foundations for the Precautionary Principle. Working Paper.

\item Thorp, E.O., 1969. Optimal gambling systems for favorable games. Review of the International Statistical Institute 37, 273-293. 

\item Thorp, E.O., 1971. Portfolio choice and the Kelly criterion. In: Proceeding

\item Ziemba, W.T. 2015 Response to Paul A Samuelson letters and papers on the Kelly capital growth investment strategy, Journal of Portfolio Management, Fall, 153-167.

\end{itemize}     

\newpage  
\clearpage

\end{document}